\documentclass[journal]{IEEEtran}
\ifCLASSINFOpdf
\else
   \usepackage[dvips]{graphicx}
\fi
\usepackage{url}
\usepackage{graphicx}
\usepackage{amsmath,bm}
\usepackage{amssymb}  
\usepackage{cite}        
 \usepackage{epstopdf} 
  \usepackage{xcolor} 
  \usepackage{subfigure} 
\usepackage{amsthm}    
\usepackage[ruled]{algorithm2e}  

\usepackage{setspace}

\begin{document}

\title{Deep Transfer Learning Based Downlink Channel Prediction for  FDD Massive MIMO Systems}

\author{Yuwen Yang,  Feifei Gao, Zhimeng Zhong,  Bo Ai,   and Ahmed Alkhateeb,
\thanks{Y. Yang and F. Gao are with  Institute for Artificial Intelligence Tsinghua University
(THUAI), State Key Lab of Intelligent Technologies and Systems, Beijing National Research Center for Information Science and
Technology (BNRist), Department of Automation, Tsinghua University, Beijing,
100084, P. R. China (email: yyw18@mails.tsinghua.edu.cn, feifeigao@ieee.org).}
\thanks{Z. Zhong is self-employed. He resides in China (e-mail: 17671455@qq.com). }
\thanks{B. Ai is with the State Key Laboratory of Rail
Traffic Control and Safety, Beijing Jiaotong University, Beijing 100044, China
(email:  boai@bjtu.edu.cn).}
\thanks{A. Alkhateeb is with the School of Electrical, Computer and
Energy Engineering at Arizona State University, Tempe, AZ 85287, USA (e-mail: alkhateeb@asu.edu). }
}

\markboth{accepted by IEEE TRANSACTIONS ON COMMUNICATIONS}
{Shell \MakeLowercase{\textit{et al.}}: xxxx}
\maketitle

\begin{abstract}
 Artificial intelligence (AI) based downlink   channel
state information (CSI)  prediction for frequency division duplexing (FDD) massive multiple-input multiple-output (MIMO) systems has  attracted growing attention recently. However, existing  works focus on  the downlink CSI prediction for the users under a given  environment  and  is hard to adapt to   users in new environment  especially when labeled data is limited. To address this issue, we formulate the downlink channel prediction  as a deep transfer learning (DTL) problem, and propose the direct-transfer algorithm based on the  fully-connected neural network  architecture, where the network is  trained in the manner of classical deep learning and is then fine-tuned  for new environments.
To further improve the transfer efficiency,  we  propose the meta-learning algorithm that
 trains the network by alternating inner-task and  across-task updates and  then  adapts to a new environment with a small number of labeled data.
 Simulation results show that the  direct-transfer  algorithm achieves better performance than  the deep learning algorithm, which implies that the transfer learning  benefits the downlink channel prediction  in new environments. Moreover, the meta-learning algorithm significantly outperforms the  direct-transfer  algorithm, which validates its effectiveness and superiority.

\end{abstract}

\begin{IEEEkeywords}
Deep transfer learning (DTL), meta-learning, few-shot learning, downlink CSI prediction, FDD, massive MIMO
\end{IEEEkeywords}

\IEEEpeerreviewmaketitle

\section{Introduction}
The acquisition of downlink  channel state information (CSI) is  a very  challenging task for frequency division duplexing (FDD) massive multiple-input multiple-output (MIMO) systems
due to the prohibitively high overheads  associated with  downlink training and uplink feedback \cite{8354789,7524027,8443598}.
By exploiting the   angular and delay reciprocities between the uplink and the downlink \cite{6328480,zhou2007experimental,hugl2002spatial},  conventional methods  proposed to
reduce the downlink training overhead by  extracting frequency-independent information from the uplink CSI, or to reduce  the uplink feedback  overhead by using compressive sensing  based algorithms
\cite{8648511,8334183,6816089,6878421}.  Nevertheless, the conventional methods either assume that the propagation paths are distinguishable and limited or highly rely on the sparsity of channels.

Recently, artificial intelligence (AI, including machine learning and deep learning,  etc.)  has been recognized as a potential  solution to deal with the high complexity and  overheads of wireless communication system \cite{8715338,8663966,8672767,8353153,8807322,8752012,8680715,8052521,8322184,8482358,guo2019convolutional,weihua20192d,alrabeiah2019viwi,alrabeiah2019deep2,Li2019,8922743,safari2018deep,8647328,alrabeiah2019deep,8795533}.
Great success has been achieved in various applications such as channel estimation \cite{8672767,8353153,8807322,8752012},  data detection \cite{8052521,8680715}, CSI feedback \cite{8322184,8482358,guo2019convolutional},  beamforming \cite{weihua20192d,alrabeiah2019viwi,alrabeiah2019deep2}, and hybrid precoding \cite{Li2019}, etc.
{Specifically, \cite{8807322} proposed to learn the  mapping from the channels of high-resolution ADC antennas to those of low-resolution ADC antennas, which achieves better performance than  existing low-resolution ADC related channel prediction methods.}
Furthermore, AI based downlink CSI prediction for FDD systems  has also been studied in  many works \cite{safari2018deep,8647328,alrabeiah2019deep,8795533}.
In \cite{safari2018deep}, a convolutional neural network is proposed to predict the downlink CSI from the uplink CSI for single-antenna FDD systems.
Based on  the  CSI correlations between different  base station (BS)  antennas, linear and  supper vector based regression methods have been proposed to use the downlink CSI of partial BS antennas to predict the whole downlink CSI, which can reduce the overheads of  both downlink pilots and uplink feedback \cite{8647328}.
In fact, \cite{alrabeiah2019deep} develops the channel mapping in space and frequent concept which proves that deep neural networks (DNNs) can learn not only the correlation between closely-located BS antennas but also between the base station arrays that are positioned at different locations or different frequency bands in the same environment.
By exploiting the mapping  relation between the uplink and downlink CSI in FDD massive MIMO systems, an efficient complex-valued based DNN  is trained  to predict the downlink CSI only from the uplink CSI, i.e., no downlink pilot is needed at all  \cite{8795533}.

Compared with conventional methods (e.g. \cite{8648511,8334183}), AI aided downlink CSI prediction benefits from
the excellent  learning capability of DNNs and  does not require  accurate channel models or high computational operations. However, existing AI aided downlink CSI prediction methods \cite{safari2018deep,8647328,8795533,alrabeiah2019deep}  all focus on training models for  users  in a certain environment.
Since  users may experience  different  wireless transmission environments,
data collection and training the models from scratch are required for the users in new environments   \cite{safari2018deep,8647328,8795533,alrabeiah2019deep}.
Typically thousands of samples and training epochs are required for training one DNN, and thus
training a new DNN for each user  would  face unacceptable time and data cost.
Therefore, it is highly desirable to design a method that can  adapt to new environments with a small amount of data.

Inspired by human's capability to  transfer knowledge from previous experience,
transfer learning, which aims to improve the performance of target tasks by exploiting the  knowledge from source tasks, becomes a promising technology in machine learning area to solve similar tasks with limited labeled data.
 Studies on  transfer learning date back from  1995  in different names, such as incremental/cumulative learning, life-long learning, multi-task learning, and learning to learn, etc \cite{5288526}.
Generally, transfer learning algorithms acquire knowledge from source tasks by pre-training a model on a large-scale source dataset and then fine-tune the pre-trained  model on a small-scale target dataset\footnote{In unsupervised learning  \cite{lin2019learning}, labeled data in target tasks are not required. Unsupervised learning algorithms require strong assumptions of  data distributions and are hard to be generalized to different problems, and therefore will not be discussed here.}.
With the rapid development of deep learning, the concept of deep transfer learning (DTL) is proposed by combining  transfer learning  with  deep learning \cite{tan2018survey}.
 The methodology in transfer learning  can be easily generalized to DTL, including instance-transfer, feature-representation-transfer, and parameter-transfer, etc \cite{5288526}.
 In particular, the model-agnostic meta-learning (MAML) algorithm proposed in \cite{finn2017model},  a classical  parameter-transfer learning algorithm,  is known  for its universal applicability and the state-of-the-art performance in few-shot learning. Unlike prior meta-learning algorithms \cite{andrychowicz2016learning,santoro2016meta,ravi2016optimization} that learn an update function and impose  restrictions on the model architecture,
the  MAML  algorithm learns a model initialization that can effectively adapt to a new task with a small amount of labeled data.


In this paper, we formulate the downlink channel prediction  for FDD massive MIMO systems as a DTL problem, where each learning task aims to predict the downlink CSI from
the uplink  CSI for  users in a certain environment.
We develop the \emph{direct-transfer  algorithm}, where the model is trained on the data from all previous environments using classical deep learning and is then fine-tuned with limited data from a new environment.
To achieve more effective transfer, we further proposed the \emph{meta-learning algorithm} that  learns a model initialization by an alternating training procedure, consisting of the  inner-task and  the across-task updates.
We also propose the \emph{no-transfer  algorithm} as a baseline algorithm, where the model is trained in the manner of classical deep learning and is directly tested on the data from a new environment without parameter adaption.
 {Furthermore, theoretical analysis and complexity comparisons are presented for the proposed three algorithms.}
Simulation results show that the  direct-transfer  algorithm outperforms the no-transfer  algorithm, which validates that transfer learning can effectively improve the performance of downlink channel prediction  in new environments. Moreover, the meta-learning algorithm significantly outperforms the  direct-transfer  algorithm,  which demonstrates its   superiority over the direct-transfer  algorithm.


The remainder of this paper is organized as follows. The
system model for FDD massive MIMO systems is introduced
in Section \ref{secmodelsdf}. The DTL problem is formulated in Section \ref{secformulation}.
The no-transfer and direct-transfer algorithms are presented  in Section \ref{secnotrans}.
The meta-learning algorithm is developed  in Section \ref{secmeta}.
Numerical results are given in Section \ref{secsimu}. Our main conclusions are
given in Section \ref{secconcul}.

\emph{Notations:}
The bold and lowercase letters denote vectors while the bold and
capital letters denote matrices.
The notations $[ \bm z ]_{p}$ and  $\textrm{len}(\bm z)$ denote the $p$-th entry and the length of the vector $\bm x$, respectively.
The notations $\Re[\cdot]$ and $\Im[\cdot]$, respectively,  denote  the real and imaginary parts of  matrices, vectors or  scales.
The notation $\left\| {\bm{x}} \right\|_1$  denotes the $L_1$  norm of  $\bm x$.
 The notation $(\cdot)^T$  denotes the transpose  of a matrix or a vector.
 The notation ${\mathbb{R}^{2M}}$ represents the $2M$-dimensional real vector space.
 The notation $ \circ $ represents the composite mapping operation.
 The notation $ \mathcal{N}_{C}(\bm{0,I})$ represents the standard
 complex Gaussian distribution.
The notation $ E [\cdot]$ represents the expectation with respect to all random variables within the brackets. The notation $\leftarrow $ represents the  assignment operation while
the notation $\rightarrow $ represents the  direction of the trajectory.

\begin{figure}[!t]
\centering
\includegraphics[width=75 mm]{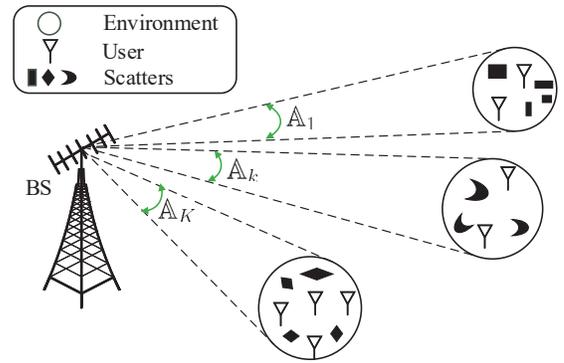}
\caption{Downlink CSI prediction for the FDD Massive MIMO system, i.e., a typical massive connectivity scenario \cite{8323218}. }
\label{figsbf}
\end{figure}

\section{System Model}\label{secmodelsdf}
We consider an
FDD massive MIMO system,
where  BS is equipped with  $M\gg 1$ antennas in the form of uniform linear array (ULA)\footnote{We adopt the ULA model here for simpler illustration, nevertheless,
the proposed approach does not restrict to the specifical array shape, and therefore is applicable for  array with arbitrary geometry.} and serves multiple  single-antenna users,   as shown in Fig.~\ref{figsbf}.
The users are randomly distributed  in $K$ regions and    users in the same region share the same propagation environment.
Denote $\mathbb{B}_k$ as the set of users in the $k$-th region.
Then, the channel between the $u$-th ($u\in \mathbb{B}_k$) user and   BS can be expressed as
\cite{8334183}:
\begin{align}\label{equedfd}
{\bm h_{u}\left(f\right)} &= \int_{\theta  \in \mathbb{A}_k} {{\alpha _u}\left( \theta  \right)}\bm a\left( \theta  \right)d\theta  \nonumber \\
&=\int_{{\bar \vartheta}_{k}^{-}}^{{\bar \vartheta}_{k}^{+}} {\left| {{\alpha _u}\left( \theta  \right)} \right|{e^{ - j2\pi {f}\tau_u \left( \theta  \right) + j\phi_u \left( \theta  \right)}}\bm a\left( \theta  \right)d\theta },
\end{align}
where ${f}$ is the carrier frequency of the $u$-th user,  while $\left| {{\alpha _u}\left( \theta  \right)} \right|$, $\phi_u \left( \theta  \right)$ and $\tau_u \left( \theta  \right)$ are the attenuation amplitude, the phase shift and the delay of the incident signal ray coming from direction of arrival (DOA) $\theta$, respectively.
Meanwhile,
$\mathbb{A}_k\buildrel \Delta \over =\left[{\bar \vartheta}_{k}^{-}, {\bar \vartheta}_{k}^{+}\right]$
is the incident angle spread (AS) of the users in the  $k$-th  region with ${\bar \vartheta}_{k}^{-}$ and ${\bar \vartheta}_{k}^{+}$ being the corresponding lower and upper bounds of AS.
The incident AS is assumed to be limited in a narrow region due to the  limited local scattering effects at the BS side
\cite{6328480,zhou2007experimental,hugl2002spatial}.
Moreover, $\bm a\left( {{\theta}} \right)$ is the the array manifold vector defined as
\begin{equation}\label{equfed}
\bm a\left( {{\theta }} \right)={\left[ {1,{e^{ - j\varpi \sin {\theta}}}, \cdots ,{e^{ - j\varpi \left( {M - 1} \right)\sin {\theta }}}} \right]^T},
\end{equation}
where $\varpi ={{2\pi d f}}/{c }$,
 $d$  is  the antenna spacing, and $c$ represents the speed of light.


\section{Formulation of DTL problem}\label{secformulation}
In this section, we  first prove the feasibility of applying deep learning to
predict the downlink CSI from uplink CSI
for a single user. Then, we formulate the   downlink channel prediction as a DTL problem and analyze the related  transfer learning algorithms.
\subsection{Deep Learning for Uplink-to-Downlink Mapping}\label{secampping}
Since the downlink and the uplink of a certain user share common
physical paths and similar spatial propagation characteristics, there exists a deterministic mapping between the downlink and the uplink channels when  the position-to-channel mapping is
bijective (see more details referring to \cite{8795533,alrabeiah2019deep}).
Denote $f_{\textrm{U},u}$ and $f_{\textrm{D},u}$ as the uplink and the downlink frequencies of the $u$-th
user, respectively. Then, the uplink-to-downlink mapping function  can be written as
\begin{eqnarray}
\bm \Psi _{f_{\textrm{U},u} \to f_{\textrm{D},u}} :\left\{ \bm h_{u}(f_{\textrm{U},u} )\right\} \to \left\{ \bm h_{u}(f_{\textrm{D},u} )\right\}.
\end{eqnarray}
Although the mapping function $\bm \Psi _{f_{\textrm{U},u} \to f_{\textrm{D},u}} $ cannot be described by known mathematical tools, it can be approximated by deep neural networks as shown in the following.

Let us consider the simplest three-layers fully-connected neural network (FNN) with only one input layer, one hidden layer with $N$  neurons,  and one output layer.
Denote $\bm x$ and $\bm \Omega$ as the input data and the network parameter of FNN, respectively. Then the corresponding output can be expressed as $\textrm{NET}_N\left( \bm x, \bm \Omega \right)$.
Since  deep learning algorithms  work in real domain, we introduce the isomorphism between the
complex and real domains as
\begin{equation}\label{euais}
\bm \xi: \bm z \to \left({\Re {\left(\bm z^T\right)},\Im {\left(\bm z^T\right)}}\right)^T.
\end{equation}
Denote the inverse mapping of $\bm \xi$ as $\bm \xi^{-1}$ that can be given as
$\bm \xi^{-1}: \left({\Re {\left(\bm z^T\right)},\Im {\left(\bm z^T\right)}}\right)^T \to \bm z$.
Based on the universal approximation theorem \cite{hornikmultilayer}, we obtain  the following proposition:

 {\emph{Proposition 1:}}
For  any given error  $\varepsilon>0$, there exists a positive
constant $N$ large enough  that satisfies
\begin{align}\label{equsdffg2}
 \mathop {\sup }\limits_{\bm x \in \mathbb{H}} \left\| {\textrm{NET}_{N}\left( \bm x, \bm \Omega \right)-\bm \Psi _{\bm \xi,u,\textrm{U} \to \textrm{D}} \left( \bm x \right)   } \right\| & \le \varepsilon, & \nonumber \\ &{\kern -60pt}  \quad \mathbb{H}=\left\{ {\bm \xi\circ \bm h_{u}(f_{\textrm{U},u} ) } \right\} \subseteq {\mathbb{R}^{2M}},
\end{align}
with
 $\bm \Psi _{\bm \xi,u,\textrm{U} \to \textrm{D}}\left( \bm x \right) = \bm \xi\circ \bm \Psi _{f_{\textrm{U},u} \to f_{\textrm{D},u}}\circ \bm \xi^{-1}\left( \bm x \right) $.
\begin{proof}
(i) Since $\forall \bm x \in \mathbb{H}$, $\left|\bm x \right|= \left|\bm  h_{u}(f_{\textrm{U},u} ) \right|\le \sqrt{M}\vartheta_u\left| {{\alpha _u}\left( \theta  \right)} \right|$, we know $\bm x $ is bounded. Therefore, $\mathbb{H}$ is a compact set;
(ii) Since $\bm \Psi _{f_{\textrm{U},u} \to f_{\textrm{D},u}}$ and $\bm \xi$ are  continuous mapping and composition of continuous mappings is still a continuous mapping,  we know  $\bm \Psi _{\bm \xi,u,\textrm{U} \to \textrm{D}} \left( \bm x \right) $ is a continuous
function for $\forall \bm x \in \mathbb{H}$. Therefore,  the $i$-th element of  $\bm \Psi _{\bm \xi,u,\textrm{U} \to \textrm{D}} \left( \bm x \right) $, denoted by $\left[\bm \Psi _{\bm \xi,u,\textrm{U} \to \textrm{D}} \left( \bm x \right) \right]_{i}$, is also a continuous function of $\bm x$.
Based on (i), (ii) and  universal approximation theorem \cite[Theorem 1]{hornikmultilayer},
we know for any $\varepsilon_i>0$, there exists a positive constant $N_i$ such that
\begin{align}
\mathop {\sup }\limits_{ x \in \mathbb{H}} \left| {\textrm{NET}_{N_i}\left(  \bm x, \bm \Omega_i \right)-\left[\bm \Psi _{\bm \xi,u,\textrm{U} \to \textrm{D}} \left(  \bm x \right)\right]_i   } \right| & \le \varepsilon_i,
\end{align}
where $\textrm{NET}_{N_i}\left(  \bm x, \bm \Omega_i \right)$ is the output of the network with only one neuron in the output layer, and $\bm \Omega_i$ denotes the corresponding network parameters.
By stacking $2M$ of the above networks together, we can construct a
larger network $\textrm{NET}_{N}\left(  \bm x, \bm \Omega\right)$ with $N=\sum_{i=1}^{2M}N_i$, where
$\bm \Omega$ denotes the corresponding network parameters. By choosing $\varepsilon_i=\varepsilon/\sqrt{2M}$, Eq.~\eqref{equsdffg2} can be proved.
\end{proof}

 {\emph{Proposition 1}} reveals that the uplink-to-downlink mapping function can be approximated
arbitrarily well by an FNN with a single hidden layer.
It should be mentioned that although the network proposed in this work cannot obtain arbitrarily high prediction accuracy due to inadequate learning or  insufficient number of hidden units \cite{hornikmultilayer},  {\emph{Proposition 1}}  still provides theoretical foundation for the application of deep neural networks.

\subsection{Definitions of DTL}
For the users in the $k$-th  environment (region), let  $\mathcal{X}_k$ and $\mathcal{Y}_k$ denote  the  space\footnote{ The  space of the uplink  channels is  $\mathcal{X}$  means that any possible uplink  channel vector  belongs to $\mathcal{X}$, i.e. $\bm h_{u}(f_{\textrm{U},u} ) \in \mathcal{X}$ holds for any possible $ u $ or $ f_{\textrm{U},u}$.} of the uplink and the downlink channels,
respectively.  Then, the definitions of the ``domain'' and the ``task'' are given in the following two definitions:

\textbf{ \emph{Definition 1:} }
 The ``domain'' $\mathcal{D}(k)$ is composed of the feature space  $\mathcal{X}_k$  and the  marginal probability distribution $P(\bm h_{u}(f_{\textrm{U},u} ))|_{u\in \mathbb{B}_k}$, i.e., $\mathcal{D}(k)=\left\{\mathcal{X}_k, P(\bm h_{u}(f_{\textrm{U},u} ))|_{u\in \mathbb{B}_k}\right\}$.

\textbf{ \emph{Definition 2:}}
 Define the ``task'' $\mathcal{T}(k)$ as the prediction of the downlink channels from the uplink channels for the users in the $k$-th environment.
  Given the specific domain $\mathcal{D}(k)$, the ``task'' $\mathcal{T}(k)$ is composed of the label space  $\mathcal{Y}_k$  and the prediction function $ \mathcal{F}_{k}$, i.e., $ \mathcal{T}(k)=\left\{\mathcal{Y}_k, \mathcal{F}_{k} \right\}$.

 \emph{Remark 1:}
 The prediction function $\mathcal{F}_{k} $ can be learned from the training data of the $k$-th  environment and then be used to predict the downlink channels for the users in the $k$-th  environment.
 Based on the analysis in Section \ref{secampping}, the prediction function $\mathcal{F}_{k} $ can  be interpreted  as the  fitting function  for all the uplink-to-downlink mapping functions defined  in the $k$-th  environment, i.e.,  $\left\{\bm \Psi _{f_{\textrm{U},u} \to f_{\textrm{D},u}}\right\}_{u\in \mathbb{B}_k}.$
 The prediction function $\mathcal{F}_{k} $ can also be interpreted as the  conditional probability  distribution $P(\bm h_{u}(f_{\textrm{D},u} )|\bm h_{u}(f_{\textrm{U},u} ))|_{u\in \mathbb{B}_k}$ from  probabilistic view.

Classical transfer learning consists of two aspects, namely, the source domain transfer and the target domain adaption.  Based on \cite{5288526}, the definition of transfer learning can be given as following:

\textbf{ \emph{Definition 3:}}
 Given the source task $\mathcal{T}_{\textrm{S}}$, the source domain $\mathcal{D}_{\textrm{S}}$,  the target task $\mathcal{T}_{\textrm{T}}$, and the target domain $\mathcal{D}_{\textrm{T}}$,
the aim of transfer learning is to improve the performance of the target task $\mathcal{T}_{\textrm{T}}$ by using the knowledge from $\mathcal{T}_{\textrm{S}}$  and $\mathcal{D}_{\textrm{S}}$, where $\mathcal{D}_{\textrm{T}}\neq \mathcal{D}_{\textrm{S}}$ or $\mathcal{T}_{\textrm{T}}\neq \mathcal{T}_{\textrm{S}} $.

Here we extend the single-source domain transfer to the multi-source domain transfer. Then, a more generalized definition of transfer learning can be provided as follows:

\textbf{ \emph{Definition 4:}}
 Given the number of source tasks $K_{\textrm{s}}$, the source tasks $\left\{\mathcal{T}_{\textrm{S}}(k)\right\}_{k=1}^{K_{\textrm{s}}}$, the source domains $\left\{\mathcal{D}_{\textrm{S}}(k)\right\}_{k=1}^{K_{\textrm{s}}}$,  the target task $\mathcal{T}_{\textrm{T}}$, and the target domain $\mathcal{D}_{\textrm{T}}$,
the aim of transfer learning is to improve the performance of the target task $\mathcal{T}_{\textrm{T}}$ by using the knowledge from $\left\{\mathcal{T}_{\textrm{S}}(k)\right\}_{k=1}^{K_{\textrm{s}}}$  and $\left\{\mathcal{D}_{\textrm{S}}(k)\right\}_{k=1}^{K_{\textrm{s}}}$, where $\mathcal{D}_{\textrm{T}}\neq \mathcal{D}_{\textrm{S}}(k)$ or $\mathcal{T}_{\textrm{T}}\neq \mathcal{T}_{\textrm{S}}(k)$ holds for $k=1,\cdots,K_{\textrm{s}}$.

In   \emph{Definition 4}, the condition $\mathcal{D}_{\textrm{T}}\neq \mathcal{D}_{\textrm{S}}(k)$ means  that either the corresponding feature space $\mathcal{X}_{\textrm{T}}\neq \mathcal{X}_{\textrm{S}}(k)$ holds or  the corresponding marginal probability distribution
 $P_{\textrm{T}}\left(\bm h_{u}(f_{\textrm{U},u} )\right)|_{u\in \mathbb{B}_{\textrm{T}}}\neq P_{\textrm{S}(k)}\left(\bm h_{u}(f_{\textrm{U},u} )\right)|_{u\in \mathbb{B}_k}$ holds, where $\mathbb{B}_{\textrm{T}}$ is the set of users in the target environment.
 The condition $\mathcal{T}_{\textrm{T}}\neq \mathcal{T}_{\textrm{S}}(k)$ means that either the label space $\mathcal{Y}_{\textrm{T}}\neq \mathcal{Y}_{\textrm{S}}(k)$ holds or the corresponding conditional probability distribution  $P_{\textrm{T}}\left(\bm h_{u}(f_{\textrm{D},u}|\bm h_{u}(f_{\textrm{U},u} ) )\right)|_{u\in \mathbb{B}_{\textrm{T}}}\neq P_{\textrm{S}(k)}\left(\bm h_{u}(f_{\textrm{D},u} )|\bm h_{u}(f_{\textrm{U},u} )\right)|_{u\in \mathbb{B}_k}$ holds. Since the conditional probability distributions for different prediction tasks are different, the condition $\mathcal{T}_{\textrm{T}}\neq \mathcal{T}_{\textrm{S}}(k)$ is satisfied.  Therefore, the downlink channel prediction for   massive MIMO systems can be formulated as  a  transfer learning problem.

DTL is to transfer knowledge by deep neural networks, which is defined as follows \cite{tan2018survey}:

\textbf{  \emph{Definition 5:}}
 Given a transfer learning task described by $\left\langle {\left\{\mathcal{T}_{\textrm{S}}(k)\right\}_{n=1}^{K_{\textrm{s}}},\left\{\mathcal{D}_{\textrm{S}}(k)\right\}_{k=1}^{K_{\textrm{s}}} \mathcal{T}_{\textrm{T}},\mathcal{D}_{\textrm{T}}} \right\rangle $, it is a DTL task when the prediction function  $\mathcal{F}_{\textrm{T}} $ of $ \mathcal{T}_{\textrm{T}}$ is a non-linear function that is approximated by a deep neural network.

Based on   \emph{Definition 4} and \emph{Definition 5}, the downlink channel prediction for   massive MIMO systems can be  formulated as a typical DTL problem, where  the $k$-th learning task is to predict the downlink channel $ \bm h_{u}(f_{\textrm{D},u} )$ from  the uplink channel $\bm h_{u}(f_{\textrm{U},u} )$ for  the
users in the $k$-th  environment.

\subsection{Motivation of Meta-learning}\label{secno}

Before resorting  to DTL methods, we should first consider the question ``whether to transfer''. In fact, if the source and target tasks are highly related,  a deep neural network
would perform well, without the need for fine-tuning the neural network based on the target environment, thanks to its remarkable generalization capability. {Therefore, a classical deep learning algorithm  should be considered as a baseline  algorithm to  investigate the necessity of transfer learning.}

The second question is ``how to transfer''.
One natural solution
is to directly use all the labeled data in the source tasks to train the network, and then  fine-tune the network with the labeled data in the target task.
However, the direct  transfer method tends to overfit when the number of labeled data in the target task is small, as will be verified by simulations in Section  \ref{secsimu}.
To overcome this challenge, and motivated by the state-of-the-art performance  of the model-agnostic meta-learning algorithm  in few-shot learning \cite{finn2017model}, we will propose  the meta-learning algorithm based downlink channel prediction  in  Section \ref{secmeta}.

\section{No-Transfer and Direct-Transfer Algorithms}\label{secnotrans}
Based on the analysis in Section \ref{secno}, we propose the no-transfer algorithm based on classical deep learning algorithms to  investigate the necessity of transfer learning. Then, the direct-transfer algorithm is proposed  and   used as the benchmark of the meta-learning algorithm. Both the no-transfer and the direct-transfer algorithms adopt the  FNN architecture as described in the following subsection.

\subsection{Network Architecture}\label{secdnaa}

\begin{figure}[!t]
\centering 
\includegraphics[width=75mm]{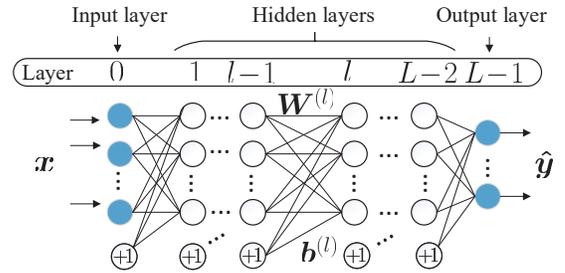}
\caption{The FNN  architecture, where the circles labeled with ``+1'' represent the bias units, blank circles represent the units in the hidden layers, and blue circles represent the units in the input and the output layers.}
\label{figdnn}       
\end{figure}

 The  FNN  architecture has  ${L}$ layers, including  one input layer,  ${L}-2$ hidden layers, and one output layer, as shown in Fig.~\ref{figdnn}.
The input of FNN is  $\bm x=\bm \xi\circ \bm h_{u}(f_{\textrm{U},u} ) $.
The output of the network is a cascade of nonlinear transformation of $\bm x$, i.e.,
\begin{equation}\label{equdnn}
\hat{\bm{y}} =  \textrm{Net}\left(\bm x, \bm\Omega \right)=\bm f^{(L-1)}\circ\bm f^{(L-2)}\circ\cdots \bm f^{(1)}\left(\bm x\right),
\end{equation}
where $\bm\Omega\buildrel \Delta \over =\left\{\bm W^{(l)},\bm b^{(l)}\right\}_{l=1}^{L-1}$  is the network parameters to be trained.
Moreover, $\bm f^{(l)}$  is the nonlinear transformation function of the  $l$-th layer and can be written as,
\begin{equation}\label{equnfd}
\bm f^{(l)}\left( \bm x \right) =\bm r^{(l)}\left(\bm W^{(l)}\bm x +\bm b^{(l)}\right),\  1\le l\le L-1,
\end{equation}
where $\bm W^{(l)}$  is the   weight matrix associated with the ($l-1$)-th and the $l$-th layers, while $\bm b^{(l)}$  and $ \bm r^{(l)}$ are the bias vector and the activation function of the  $l$-th layer, respectively. The activation function for the hidden layers  is selected as
  the rectified linear unit (ReLU) function
   $\left[\bm r_{\textrm{re}}(\bm z)\right]_{p}=\max\{0,\left[\bm z\right]_{p}\}$ with $\ p= 1,2,\cdots,\textrm{len}(\bm z)$.  The activation function
for the  output layer is  the linear function, i.e.,  $\bm r_{\textrm{li}} (\bm z) =\bm z$.
The loss  function can be written as follow:
\begin{equation}\label{equloss}
\textrm{Loss}_{\mathbb{D}}\left(\bm \Omega\right) =\frac{1}{{V}}\sum\limits_{v= 0}^{ V-1}\left\|\hat{\bm{y}}^{(v)}-{\bm{y}}^{(v)} \right\|_{2}^{2},
\end{equation}
where  $\bm y=\bm \xi \circ \bm h_{u}(f_{\textrm{D},u} ) $ is the supervise label,
 $V$ is the batch size\footnote{Batch size is the number of samples in one training batch.}, the superscript $(v)$ denotes
the index of the $v$-th  training sample, ${\mathbb{D}}=\left\{(\bm x^{(v)},\bm y^{(v)})\right\}_{v=1}^{V} $ is the  training data of one batch, and ${\left\|{\cdot}\right\|}_{2}$ denotes the  $L_2$ norm.

\subsection{Definitions and Generation of Datasets}\label{secdata}

As mentioned in   \emph{Definition 2}, each task represents the prediction of downlink channel from the uplink channel for  the users in a certain environment.
In the $k$-th environment, $U$  different users are involved in the  generation of datasets.
 Denote $\Delta f$ as the frequency difference between the uplink and the downlink. Then, each uplink frequency corresponds to a certain downlink frequency,  i.e., $f_{\textrm{D},u}=f_{\textrm{U},u}+\Delta f$.
 Denote the  training dataset of the   $k$-th source task as $\mathbb{D}_{\textrm{Tr}}(k)$.
We can collect $N_{\textrm{Tr}}$ sample pairs $ \left\{ \left(\bm \xi\circ\bm h_{u}(f_{\textrm{U},u} ), \bm \xi\circ\bm h_{u}(f_{\textrm{D},u} )\right)\right\}$ as  the dataset $\mathbb{D}_{\textrm{Tr}}(k)$ by  randomly selecting multiple uplink frequencies  for  users in the $k$-th environment.


Denote the  adaption and testing  datasets of the $k$-th  target task as $\mathbb{D}_{\textrm{Ad}}(k)$ and  $\mathbb{D}_{\textrm{Te}}(k)$, respectively.
Similarly, the datasets  $\mathbb{D}_{\textrm{Ad}}(k)$ and  $\mathbb{D}_{\textrm{Te}}(k)$ can be obtained by  separately collecting $N_{\textrm{Ad}}$ and $N_{\textrm{Te}}$ sample pairs for the $k$-th target task.
Note that $\mathbb{D}_{\textrm{Ad}} (k)\cap \mathbb{D}_{\textrm{Te}} (k)= \emptyset $ should be satisfied to ensure the testing dataset  not known to the networks.

\subsection{No-Transfer Algorithm}\label{secnot}
The no-transfer algorithm regards the DTL problem as one classical deep learning problem,  including the training  and the testing stages.

In the training stage, the  datasets  $\left\{\mathbb{D}_{\textrm{Tr}}(k)\right\}_{k=1}^{K_{\rm{S}}}$ for $K_{\rm{S}}$ source tasks are collected as a  complete training dataset $\mathbb{D}_{\textrm{STr}}$.
 In each time step, $V$ training samples are randomly selected from $\mathbb{D}_{\textrm{STr}}$ as the training batch  $\mathbb{D}_{\textrm{STrB}}$.
 Then, we employ the adaptive moment estimation (ADAM) algorithm \cite{kingmaadam} to minimize the loss function on  $\mathbb{D}_{\textrm{STrB}}$, i.e., $\textrm{Loss}_{\mathbb{D}_{\textrm{STrB}}}\left( \bm \Omega  \right)$.
 Instead of using the gradient of the current step to guide the updates, ADAM adopts  moments of the gradient  to determine the direction of the optimization,  which can accelerate the training, especially in the case of  high curvature or noisy gradients
  \cite[chapter 8]{goodfellow2016deep}.
By iteratively executing the  ADAM algorithm on the  dataset $\mathbb{D}_{\textrm{STr}}$ until  $\textrm{Loss}_{\mathbb{D}_{\textrm{STrB}}}\left( \bm \Omega  \right)$ converges, we can obtain the trained network parameter  $\bm \Omega_{\rm{Nt}}$.

 In the testing stage, the network parameter  $\bm \Omega_{\rm{Nt}}$ is fixed.
 The  testing datasets $\left\{\mathbb{D}_{\textrm{Te}}(k)\right\}_{k=1}^{K_{\rm{T}}}$  of $K_{\rm{T}}$ target tasks  are generated and are used to  test the performance of the no-transfer algorithm.
Denote $\hat{\bm h}_{\textrm{D}}=\bm \xi^{-1} \circ\hat{\bm y}$ and  ${\bm h}_{\textrm{D}}=\bm \xi^{-1}\circ{\textbf{}\bm y} $ as the estimated and the true downlink channel, respectively.
 Normalized mean-squared-error (NMSE) is used to measure the prediction accuracy, which is defined as
\begin{equation}\label{equddmse}
{\rm{NMSE}} = E \left[\left\|{\bm h}_{\textrm{D}}-\hat{\bm h}_{\textrm{D}}\right\|_{2}^{2}/\left\|{\bm h}_{\textrm{D}}\right\|_{2}^{2}\right].
\end{equation}
Denote the prediction NMSE of the no-transfer algorithm
 as $\rm{NMSE}_{\textrm{Nt}}$ that  can be calculated by  averaging the NMSEs of the no-transfer algorithm  evaluated on $ {K_{\textrm{T}}}$ target environments. i.e., $ \rm{NMSE}_{\textrm{Nt}}=\sum_{k=1}^{K_{\textrm{T}}}\rm{NMSE}_{\textrm{Nt}}(k)/{K_{\textrm{T}}}
$,  where
 $\rm{NMSE}_{\textrm{Nt}}(k)$ is the NMSE  evaluated on the $k$-th target environment.
The concrete steps of  the no-transfer algorithm are given in the   \textbf{Algorithm~1}.

\begin{algorithm}
    \LinesNumbered
    \caption{No-transfer algorithm for  downlink CSI prediction}
    \label{notr}
    \KwIn{ Source tasks: $\left\{\mathcal{T}_{\textrm{S}}(k)\right\}_{k=1}^{K_{\textrm{S}}}$, Target tasks: $\left\{\mathcal{T}_{\textrm{T}}(k)\right\}_{k=1}^{K_{\textrm{T}}}$, learning rate: $\gamma$,
     batch size: $V$}
     \KwOut {Trained network parameter: $\bm \Omega_{\textrm{Nt}}\leftarrow\bm \Omega$, predicted downlink CSI based on $\left\{\mathbb{D}_{\textrm{Te}}(k)\right\}_{k=1}^{K_{\textrm{T}}}$, NMSE of the no-transfer algorithm: $\rm{NMSE}_{\textrm{Nt}}$}
     \texttt{Training stage}
    \\ Randomly initialize the network parameters $\bm \Omega$ \label{trssin}
    \\ Generate the  training dataset $\mathbb{D}_{\textrm{STr}}\buildrel \Delta \over = \left\{\mathbb{D}_{\textrm{Tr}}(k)\right\}_{k=1}^{K_{\rm{S}}}$ for $K_{\rm{S}}$ source tasks\\
   \For {$t=1,\cdots$} {\label{trr}
        Randomly select $V$ training samples from $\mathbb{D}_{\textrm{STr}}$ as the training batch  $\mathbb{D}_{\textrm{STrB}}$
       \\   {Update  $\bm \Omega$ by using the  ADAM algorithm (learning rate $\gamma$) to minimize $\textrm{Loss}_{\mathbb{D}_{\textrm{STrB}}}\left( \bm \Omega  \right)$}
             }
 \texttt{Testing stage}\\
 Initialize  NMSE: $\rm{NMSE}_{\textrm{Nt}}  \leftarrow 0$\\
    \For{$k=1,\cdots,K_{\rm{T}}$}{
        Generate the testing dataset $\mathbb{D}_{\textrm{Te}}(k)$ for  $\mathcal{T}_{\textrm{T}} (k)$
       \\  Predict the downlink CSI base on  $\mathbb{D}_{\textrm{Te}}(k)$ and $\bm \Omega$ using Eq.~\eqref{equdnn}
       \\  Calculate $\rm{NMSE}_{\textrm{Nt}} (k)$  using Eq.~\eqref{equddmse}
        \\ $\rm{NMSE}_{\textrm{Nt}} \leftarrow \rm{NMSE}_{\textrm{Nt}} +\rm{NMSE}_{\textrm{Nt}} (k)/K_{\rm{T}}$
        }
\end{algorithm}
\subsection{Direct-transfer Algorithm}
In the training stage, the  direct-transfer algorithm  trains the network via the ADAM algorithm on the training dataset $\mathbb{D}_{\textrm{STr}}$ until  $\textrm{Loss}_{\mathbb{D}_{\textrm{STrB}}}\left( \bm \Omega  \right)$ converges.
During  the training, the network tries to learn a  generalized  parameter to predict the downlink CSI for  users in different environments.
After the training finished, the network has learned the  parameter $\bm\Omega_{\textrm{Nt}}$ from the $K_{\textrm{S}}$ source tasks.
Then, the  datasets $\left\{\mathbb{D}_{\textrm{Ad}}(k)\right\}_{k=1}^{k=K_{\textrm{T}}}$ and  $\left\{\mathbb{D}_{\textrm{Te}}(k)\right\}_{k=1}^{k=K_{\textrm{T}}}$ of  $K_{\textrm{T}}$ target tasks are generated  following Section \ref{secdata}.
For the $k$-th target task, the direct-transfer algorithm
 initializes the target-task-specific parameter $\bm \Omega_{\textrm{T},k}$ as  the trained network parameter $\bm \Omega_{\textrm{Nt}}$, then fine-tunes $\bm \Omega_{\textrm{T},k}$ on the adaption dataset $\mathbb{D}_{\textrm{Ad}}(k)$ via  $G_{\textrm{Ad}}$  steps of ADAM updates.
After the fine-tuning  finished, the target-task-specific parameter $\bm \Omega_{\textrm{T},k}$ will be fixed. Let $\rm{NMSE}_{\textrm{Dt}}(k)$ represent the NMSE of the direct-transfer algorithm evaluated on the $k$-th target environment that can be obtained by testing the network on the dataset $\mathbb{D}_{\textrm{Te}}(k)$ using  Eqs.~\eqref{equdnn} and \eqref{equddmse}.
Then, the  prediction NMSE of the direct-transfer algorithm can be obtained by
 $\rm{NMSE}_{\textrm{Dt}}\!=\!\sum_{k=1}^{K_{\textrm{T}}}\rm{NMSE}_{\textrm{Dt}}(k)/{K_{\textrm{T}}}
$. The concrete steps of  the direct-transfer algorithm are given in the   \textbf{Algorithm~2}.

\begin{algorithm}
    \caption{Direct-transfer algorithm for   downlink CSI prediction}
    \KwIn{ Source tasks: $\left\{\mathcal{T}_{\textrm{S}}(k)\right\}_{k=1}^{K_{\textrm{S}}}$, Target tasks: $\left\{\mathcal{T}_{\textrm{T}}(k)\right\}_{k=1}^{K_{\textrm{T}}}$, learning rate: $\beta$, batch size: $V$, number of gradsteps for adaption: $G_{\textrm{Ad}}$}
    \KwOut{Trained network parameter: $\bm \Omega_{\textrm{Nt}}$, predicted downlink CSI based on   $\left\{\mathbb{D}_{\textrm{Te}}(k)\right\}_{k=1}^{K_{\textrm{T}}}$, NMSE of the direct-transfer algorithm: $\rm{NMSE}_{\textrm{Dt}}$}
\texttt{Training stage}
   \\  Implement the  \textbf{Algorithm~1} and obtain the trained network parameter $ \bm \Omega_{\textrm{Nt}}$
   \\ \texttt{Direct-adaption and Testing}
\\ Initialize  NMSE: $\rm{NMSE}_{\textrm{Dt}}  \leftarrow 0$\\
    \For{$k=1,\cdots,K_{\rm{T}}$}{
        Generate the  datasets $\mathbb{D}_{\textrm{Ad}}(k)$ and  $\mathbb{D}_{\textrm{Te}}(k)$ for   $\mathcal{T}_{\textrm{T}} (k)$
          \\  \texttt{Direct-adaption stage}
        \\ Load the network parameter $\bm \Omega_{\textrm{T},k} \leftarrow \bm \Omega_{\textrm{Nt}}$\\
        \For{$g=1,\cdots,G_{\rm{Ad}}$}{
        {Update  $\bm \Omega_{\textrm{T},k} $ by using   ADAM  (learning rate $\beta$) to minimize $\textrm{Loss}_{\mathbb{D}_{\textrm{Ad}}(k)}\left( \bm \Omega_{\textrm{T},k}\right)$}
         }
          \texttt{Testing stage}
        \\ Predict the downlink CSI base on  $\mathbb{D}_{\textrm{Te}}(k)$ and $\bm \Omega_{\textrm{T},k} $ using Eq.~\eqref{equdnn}
        \\ Calculate $\rm{NMSE}_{\textrm{Dt}} (k)$  using Eq.~\eqref{equddmse}
        \\ $\rm{NMSE}_{\textrm{Dt}} \leftarrow \rm{NMSE}_{\textrm{Dt}} +\rm{NMSE}_{\textrm{Dt}} (k)/K_{\rm{T}}$}
\end{algorithm}

\begin{figure}[!t]
\centering
\includegraphics[width=75mm]{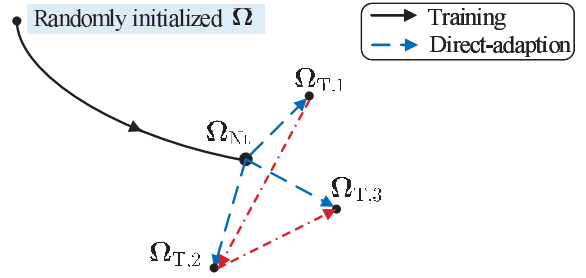}
\caption{Illustration of the direct-transfer algorithm, where $K_{\textrm{T}}=3$.}
\label{figinitial}
\end{figure}

Notice that the direct-transfer algorithm utilizes the trained  network parameter $ \bm \Omega_{\textrm{Nt}}$ as the  initialization of the direct-adaption stage for every target task.
By contrast, another way is to utilize previous target-task-specific parameter  $\bm \Omega_{\textrm{T},k-1}$ as the  initialization of the  direct-adaption stage for the $k$-th target task
since  $\bm \Omega_{\textrm{T},k-1}$ appears  to contain the information  in all the source environments and $k-1$ target environments, and therefore could enhance the performance of the algorithm.
However,  the real situation is that  a good initialization in DTL  is not to train the network parameter with as much data as possible, but to find  such  a parameter that can
 easily  adapt to any target task. In a way,
a good initial parameter  can be interpreted as a parameter point in the parameter space that is near (or easy to get) to the optimal parameter point  for  most target tasks.
Fig.~\ref{figinitial} illustrates the trajectory of parameter updating during the training and  direct-adaption  stages of the direct-transfer algorithm.
The  black solid line represents the updating trajectory of the network parameter $\bm \Omega$ during the training stage. The blue dashed lines represent the updating trajectories of the target-task-specific parameters $\left\{\bm \Omega_{\textrm{T},k}\right\}_{k=1}^{3}$  during the direct-adaption stage following \textbf{Algorithm~2}. The
red dash-dotted lines represent  the parameter adaption process with   previous target-task-specific parameter as the initial parameter.
As shown in Fig.~\ref{figinitial}, the trajectory of red dash-dotted lines, i.e., $\bm \Omega_{\textrm{Nt}} \rightarrow\bm \Omega_{\textrm{T},1}\rightarrow\bm \Omega_{\textrm{T},2}\rightarrow\bm \Omega_{\textrm{T},3}$,
is longer than the trajectory of blue dashed lines, i.e., $\left\{\bm \Omega_{\textrm{Nt}} \rightarrow\bm \Omega_{\textrm{T},k}\right\}_{k=1}^{3}$, which indicates $\bm \Omega_{\textrm{Nt}} $ is a better initialization than  $\bm \Omega_{\textrm{T},k-1}$. The intrinsic reason is that $\bm \Omega_{\textrm{Nt}} $ is obtained by training the network on a large-scale and shuffled dataset, i.e., $\mathbb{D}_{\textrm{STr}}$,
which has more stronger representativeness of different tasks than the small-scale adaption dataset $\mathbb{D}_{\textrm{Ad}}(k)$. In fact, the fine-tuning on the adaption dataset $\mathbb{D}_{\textrm{Ad}}(k)$ renders the network parameter away from the generalized parameter $\bm \Omega_{\textrm{Nt}} $ and gets closed to the task-specific parameter $\bm \Omega_{\textrm{T},k}$.

\emph{Remark 2:}
It should be emphasized that parameter initialization is one of the most crucial tasks in deep learning, especially in DTL. A high-quality  initialization can effectively prevent the training from getting stuck in a low-performance local minimum and thus  accelerates the
 convergence \cite{liangunderstanding,sutskeverimportance}.

\section{Meta-learning Algorithm}\label{secmeta}
The proposed meta-learning algorithm also adopts the FNN architecture   in  Section \ref{secdnaa} and  has three stages,  namely, the meta-training, the  meta-adaption  and the testing stages.

\subsection{Definitions and Generation of Datasets}\label{secdatadsd}
For the $k$-th source task, two datasets should be obtained for the meta-training stage, i.e., the training support dataset $\mathbb{D}_{\textrm{TrSup}}(k)$ and the training query dataset  $\mathbb{D}_{\textrm{TrQue}}(k)$.
Following  Section~\ref{secdata}, we can  collect multiple sample pairs for each source task and then  randomly divide these sample pairs, i.e.,   $ \left\{ \left(\bm \xi\circ\bm h_{u}(f_{\textrm{U},u} ), \bm \xi\circ\bm h_{u}(f_{\textrm{D},u} )\right)\right\}$, into the   training support dataset $\mathbb{D}_{\textrm{TrSup}}(k)$ and the training query dataset  $\mathbb{D}_{\textrm{TrQue}}(k)$.
  Note that  $\mathbb{D}_{\textrm{TrSup}}(k) \cap \mathbb{D}_{\textrm{TrQue}} (k)= \emptyset $
  should be satisfied to improve the generalization  capability of the network.
For the $k$-th target task, the  adaption  dataset $\mathbb{D}_{\textrm{Ad}}(k)$ and the
   testing dataset  $\mathbb{D}_{\textrm{Te}}(k)$ can be obtained following Section~\ref{secdata}.

\subsection{Meta-training Stage}
In the meta-training stage, the  meta-learning algorithm aims to learn  a network initialization
 that  can effectively  adapt to a new task. The parameter of FNN, i.e., $\bm \Omega$, is  randomly initialized and is then  updated by  two  iterative  processes, i.e., the inner-task  and the across-task updates.

\subsubsection{Inner-task update}
Denote the source-task-specific parameter as $\bm \Omega_{\textrm{S},k}$.
The goal of the inner-task updates is to minimize the loss function on the training support dataset $\mathbb{D}_{\textrm{TrSup}}(k)$, i.e.,  $\textrm{Loss}_{\mathbb{D}_{\textrm{TrSup}}(k)}\left( \bm \Omega_{\textrm{S},k}\right)$,  by iteratively updating the parameter $\bm \Omega_{\textrm{S},k}$. Specifically,
  $\bm \Omega_{\textrm{S},k}$ is  initialized as the current network parameter $\bm \Omega$, and is then updated with $G_{\rm{Tr}}$ steps of gradient descents, i.e.,
 \begin{eqnarray}\label{equupda34}
\bm \Omega_{\textrm{S},k} \leftarrow\bm \Omega_{\textrm{S},k}-\beta {\nabla_{\bm \Omega_{\textrm{S},k}} }\textrm{Loss}_{\mathbb{D}_{\textrm{TrSup}}(k)}\left( \bm \Omega_{\textrm{S},k}\right),
 \end{eqnarray}
where $\beta$ is the inner-task learning rate.

\subsubsection{Across-task update}

In the $t$-th time step, we randomly select
$K_{\rm{B}}$ source tasks  out of  $K_{\rm{S}}$ source tasks\footnote{Typically, there is $K_{\rm{B}}\ll K_{\rm{S}}$.} and generate corresponding support datasets $\left\{\mathbb{D}_{\textrm{TrSup}}(k)\right\}_{k=1}^{K_{\textrm{B}}}$ and query datasets $\left\{\mathbb{D}_{\textrm{TrQue}}(k)\right\}_{k=1}^{K_{\textrm{B}}}$ following Section~\ref{secdatadsd}.
The optimization objective of  the $t$-th update is the loss function associated with the $K_{\rm{B}}$ training query datasets $\left\{\mathbb{D}_{\textrm{TrQue}}(k)\right\}_{k=1}^{K_{\textrm{B}}}$ and the meta-trained source-task-specific parameter $\bm \Omega_{\textrm{S},k}$, i.e.,
\[\sum\limits_{k=1}^{K_{\rm{B}}}\textrm{Loss}_{\mathbb{D}_{\textrm{TrQue}}(k)}\left( \bm \Omega_{\textrm{S},k}  \right).\]
The network parameter  $\bm \Omega$ is updated  via the ADAM algorithm with learning rate $\gamma$.
After the across-task updates finished, the  meta-trained network parameter will be obtained by $\bm \Omega_{\textrm{Mt}}\leftarrow  \bm \Omega$.
By alternately implementing the inner-task and across-task updates until the loss  converges, the network learns model initialization that can
 adapts to a new task using only a small number of
samples.

There are two important differences between the inner-task and  across-task updates.
(i)
 Each inner-task update is performed on the support dataset of one  source task while
 each across-task update  is performed on the query datasets of $K_{\rm{B}}$ source tasks;
(ii) The aim of inner-task updates is to optimize the task-specific parameters $\left\{\bm \Omega_{\textrm{S},k}\right\}_{k=1}^{K_{\rm{B}}}$ while the aim of across-task updates is to optimize the overall network  parameters $\bm \Omega$.


\subsection{Meta-adaption and Testing Stages}
Following Section \ref{secdatadsd}, we generate the adaption datasets $\left\{\mathbb{D}_{\textrm{Ad}}(k)\right\}_{k-1}^{k=K_{\textrm{T}}}$ and  the  testing datasets $\left\{\mathbb{D}_{\textrm{Te}}(k)\right\}_{k-1}^{k=K_{\textrm{T}}}$   for the meta-adaption and testing stages, respectively.
In the meta-adaption stage, the  meta-learning  algorithm aims to fine-tune the network  on the adaption dataset $\mathbb{D}_{\textrm{Ad}}(k)$  so that the network can predict the  downlink CSI  on $\mathbb{D}_{\textrm{Te}}(k)$ as  accurate as possible.
For the $k$-th target task, the direct-transfer algorithm
 initializes the target-task-specific parameter $\bm \Omega_{\textrm{T},k}$ as  the meta-trained network parameter $\bm \Omega_{\textrm{Mt}}$, and then fine-tunes $\bm \Omega_{\textrm{T},k}$ on the adaption dataset $\mathbb{D}_{\textrm{Ad}}(k)$ via  $G_{\textrm{Ad}}$  steps of gradient descents.
The optimization objective is the loss function on the $\mathbb{D}_{\textrm{Ad}}(k)$, i.e.,
$\textrm{Loss}_{\mathbb{D}_{\textrm{Ad}}(k)}\left( \bm \Omega_{\textrm{T},k}\right)$.
Therefore, the  parameter $\bm \Omega_{\textrm{T},k}$ can be updated  by
 \begin{eqnarray}\label{equupdatsd2}
\bm \Omega_{\textrm{T},k} \leftarrow\bm \Omega_{\textrm{T},k}-\beta {\nabla _{\bm \Omega_{\textrm{T},k}} }\textrm{Loss}_{\mathbb{D}_{\textrm{Ad}}(k)}\left( \bm \Omega_{\textrm{T},k}\right).
 \end{eqnarray}
After the fine-tuning  finished, the target-task-specific parameter $\bm \Omega_{\textrm{T},k}$ will be fixed. Let $\rm{NMSE}_{\textrm{Mt}}(k)$ represent the NMSE of the  meta-learning  algorithm evaluated on the $k$-th target environment that can be obtained by testing the network on the dataset $\mathbb{D}_{\textrm{Te}}(k)$ using  Eqs.~\eqref{equdnn} and \eqref{equddmse}.
Then, the  prediction NMSE of the meta-learning  algorithm  can be obtained by
 $\rm{NMSE}_{\textrm{Mt}}\!=\!\sum_{k=1}^{K_{\textrm{T}}}\rm{NMSE}_{\textrm{Mt}}(k)/{K_{\textrm{T}}}
$. The concrete steps of  the meta-learning algorithm are given in    \textbf{Algorithm~3}.

\begin{algorithm}
    \LinesNumbered
\label{algo1}
    \caption{ Meta-learning algorithm for  downlink CSI prediction}
    \KwIn { Source tasks: $\left\{\mathcal{T}_{\textrm{S}}(k)\right\}_{k=1}^{K_{\textrm{S}}}$, Target tasks: $\left\{\mathcal{T}_{\textrm{T}}(k)\right\}_{k=1}^{K_{\textrm{T}}}$, inner-task learning rate: $\beta$, across-task learning rate: $\gamma$,
    number of  gradsteps for inner-task training: $G_{\textrm{Tr}}$, number of tasks in each time step: $K_{\rm{B}}$}
    \KwOut {Meta-trained network parameter: $\bm \Omega_{\textrm{Mt}}$, Predicted downlink CSI based on   $\left\{\mathbb{D}_{\textrm{Te}}(k)\right\}_{k=1}^{K_{\textrm{T}}}$, NMSE of the meta-learning algorithm: $\rm{NMSE}_{\textrm{Mt}}$ }
 \texttt{Meta-training stage}
    \\ Randomly initialize the network parameters $\bm \Omega$\\
    \For{$t=1,\cdots$}{
         Randomly select $K_{\rm{B}}$ tasks from $\left\{\mathcal{T}_{\textrm{S}}(k)\right\}_{k=1}^{K_{\textrm{S}}}$\\
      Generate corresponding   datasets $\left\{\mathbb{D}_{\textrm{TrSup}}(k)\right\}_{k=1}^{K_{\textrm{B}}}$ and $\left\{\mathbb{D}_{\textrm{TrQue}}(k)\right\}_{k=1}^{K_{\textrm{B}}}$ \\
        \For{$k=1,\cdots,K_{\rm{B}}$}{
         Initialize the parameter $\bm \Omega_{\textrm{S},k} \leftarrow \bm \Omega$ \label{inner}\\
            \For{$g=1,\cdots,G_{\rm{Tr}}$}{
             Update  $\bm \Omega_{\textrm{S},k}$ using  Eq.~\eqref{equupda34}}
        Update $\bm \Omega $ by using ADAM (learning rate $\gamma$) to minimize
        $\sum_{k=1}^{K_{\rm{B}}}\textrm{Loss}_{\mathbb{D}_{\textrm{TrQue}}(k)}\left( \bm \Omega_{\textrm{S},k}  \right)$ }}
      Meta-trained network parameter: $\bm \Omega_{\textrm{Mt}}\leftarrow \bm \Omega $\\
     \texttt{Meta-adaption and Testing}\\
  Initialize  NMSE: $\rm{NMSE}_{\textrm{Mt}}  \leftarrow 0$\\
    \For{$k=1,\cdots,K_{\rm{T}}$}
    {
        Generate the  datasets $\mathbb{D}_{\textrm{Ad}}(k)$ and  $\mathbb{D}_{\textrm{Te}}(k)$ for   $\mathcal{T}_{\textrm{T}} (k)$ \\
          \texttt{Meta-adaption stage}\\
       Load the network parameter $\bm \Omega_{\textrm{T},k} \leftarrow \bm \Omega_{\textrm{Mt}}$\\
        \For{$g=1,\cdots,G_{\rm{Ad}}$}
        {
         Update  $\bm \Omega_{\textrm{T},k}$ using  Eq.~\eqref{equupdatsd2}
        }
        \texttt{Testing stage} \\
        Predict the downlink CSI base on  $\mathbb{D}_{\textrm{Te}}(k)$ and $\bm \Omega_{\textrm{T},k} $ using Eq.~\eqref{equdnn}\\
      Calculate $\rm{NMSE}_{\textrm{Mt}} (k)$  using Eq.~\eqref{equddmse}\\
      $\rm{NMSE}_{\textrm{Mt}} \leftarrow \rm{NMSE}_{\textrm{Mt}} +\rm{NMSE}_{\textrm{Mt}} (k)/K_{\rm{T}}$
    }
\end{algorithm}

 {\emph{Remark 3:}
The proposed meta-learning algorithm is inspired by the MAML algorithm  in \cite{finn2017model}. While it should be mentioned that several  modifications have been made in the proposed meta-learning algorithm:}

 {i) The inner-task update only involves one step of gradient descent in  \cite{finn2017model} while the proposed algorithm adopts $G_{\rm{Tr}}$  steps to obtain better performance;}


\begin{figure}[!t]
\centering
\includegraphics[width=75 mm]{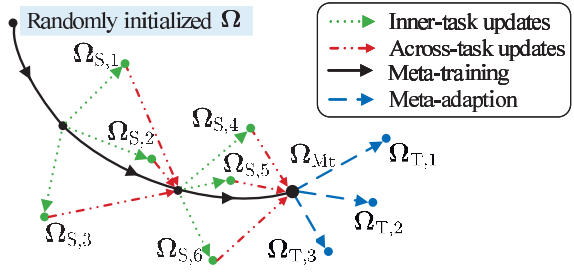}
\caption{Illustration of the meta-learning algorithm, where $K_{\textrm{B}}=K_{\textrm{T}}=3$.}
\label{figinmeta}
\end{figure}

 {ii) The across-task update  is preformed by the stochastic gradient descent (SGD) algorithm
in  \cite{finn2017model} while  the proposed algorithm adopts the ADAM algorithm instead of the SGD algorithm. This is because that  the ADAM benefits from the moment technique and the parameter-specific learning rates, and thus  outperforms SGD in many situations \cite{chen2019decaying}.
}

Fig.~\ref{figinmeta} illustrates  the trajectories of parameter updating during the meta-training and meta-adaption stages of the  meta-learning algorithm.
The  black solid line represents the updating trajectory of the network parameter $\bm \Omega$ during the meta-training stage.
The blue dashed lines represent the updating trajectories of the target-task-specific parameters $\left\{\bm \Omega_{\textrm{T},k}\right\}_{k=1}^{3}$  during the meta-adaption stage.
The green dotted lines represent the  trajectories of the source-task-specific parameters $\left\{\bm \Omega_{\textrm{S},k}\right\}_{k=1}^{6}$  during the inner-task updates.
As shown in Fig.~\ref{figinmeta}, $K_{\textrm{B}}$ red dash-dotted lines point to the next position of parameter $\bm \Omega$, which implies that $K_{\textrm{B}}$ trained source-task-specific parameters are involved in the across-task update to determine the updating direction of $\bm \Omega$.
As depicted in Fig.~\ref{figinitial} and Fig.~\ref{figinmeta}, the  meta-learning algorithm takes the  unique features of different tasks into account and  exploits the joint guidance of multiple source tasks while the direct-transfer algorithm simply regards all the   source tasks as one source task and ignores the structure information in different tasks.

 {\subsection{Theoretical Analysis of the Meta-learning Algorithm}
To understand why the meta-learning algorithm outperforms the direct-transfer algorithm, we focus on the loss function that is used to guide the updates of $\bm \Omega$ for the meta-learning algorithm, i.e., $\sum_{k=1}^{K_{\rm{B}}}\textrm{Loss}_{\mathbb{D}_{\textrm{TrQue}}(k)}\left( \bm \Omega_{\textrm{S},k}  \right)$.
We add the  superscript $(t)$ to distinguish the parameters at different time steps and set $G_{\rm{Tr}}$  as 1 to simplify the parameter updates in the meta-training stage.
Based on \textbf{Algorithm~3},  the  loss function at the $t$-th time step is computed as
\begin{subequations}
\begin{align}
& {\kern 10pt} \sum\limits_{k=1}^{K_{\rm{B}}}\textrm{Loss}_{\mathbb{D}_{\textrm{TrQue}}(k)}\left( \bm \Omega_{\textrm{S},k}^{(t)}  \right)   \\
 &=  \sum\limits_{k=1}^{K_{\rm{B}}}\textrm{Loss}_{\mathbb{D}_{\textrm{TrQue}}(k)}\left( \bm \Omega^{(t-1)}-\beta \nabla \textrm{Loss}_{\mathbb{D}_{\textrm{TrSup}}(k)}\left( \bm \Omega^{(t-1)}  \right)\right)  \nonumber\\
  &\approx    \sum\limits_{k=1}^{K_{\rm{B}}}\textrm{Loss}_{\mathbb{D}_{\textrm{TrQue}}(k)}\left( \bm \Omega^{(t-1)}\right) \nonumber \\  &{\kern 10pt} -  \sum\limits_{k=1}^{K_{\rm{B}}}\beta \nabla \textrm{Loss}_{\mathbb{D}_{\textrm{TrSup}}(k)}\left( \bm \Omega^{(t-1)}  \right) \nabla \textrm{Loss}_{\mathbb{D}_{\textrm{TrQue}}(k)}\left( \bm \Omega^{(t-1)}  \right), \label{subequ}
\end{align}
\end{subequations}
where Eq.~\eqref{subequ} is obtained based on  the first-order Taylor expansion.
The first term  in  Eq.~\eqref{subequ} represents the
 the sum of the losses on query datasets.
 The second term  in  Eq.~\eqref{subequ} represents the sum of
the negative inner products of the gradients  on the  query and the support datasets.
We know that if  the directions of the two gradients are closer, then
 the negative inner product will be smaller, which indicates that during the meta-training stage,
 the algorithm tries to maximize the similarity between the gradients on the  query and the support datasets.  Compared with the no-transfer and the direct-transfer algorithms, the meta-learning algorithm enhances the  generalization capability  between the  query and the support datasets (also  between  the adaption and testing datasets), and therefore can adapt to  a new task more effectively \cite{Alex22}.}

\section{Simulation Results}\label{secsimu}
In this section, we will first present the simulation scenario and default algorithm parameters.
Then, the  performance of the no-transfer, the direct-transfer, and the meta-learning algorithms
will be evaluated and analyzed.

\subsection{Simulation Setup}
\begin{figure*}[!t]
\centering
\includegraphics[width=120mm]{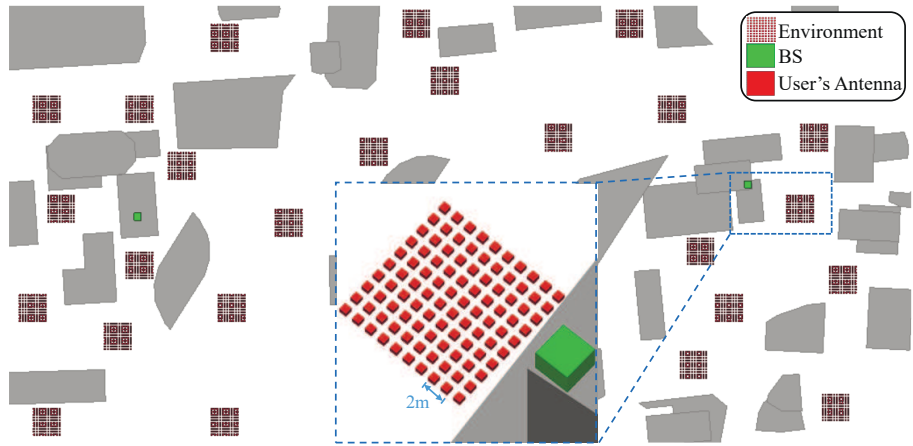}
\caption{A partial view of the ray-tracing scenario. The green little box represents the BS antennas.  The red little box represents the user antennas.   The red  square array represents the user antennas in the same environments, where the distance between adjacent red little boxes is 2 m.  T {his ray-tracing simulator  shoots thousands of rays in all directions from the transmitter and records the strongest 25 paths that reach the receiver \cite{timmurphy}.}}
\label{figtop}
\end{figure*}

In the simulations, we consider the outdoor massive MIMO scenario that is constructed based on   the accurate 3D ray-tracing simulator  \cite{timmurphy}.
The scenario comprises  4 BSs  and massive randomly distributed  user antennas and each BS is  equipped with 64 antennas. The scenario covers an area of $1500\times1500$ square metres.
A partial view of the ray-tracing scenario is illustrated in  Fig.~\ref{figtop}.
There are total 1300 environments randomly distributed in the outdoor massive MIMO scenario, where each environment contains multiple possible user locations as shown in the  partial enlarged picture.

To generate the training dataset, the uplink frequency is randomly selected in $[1,3]$ GHz. The frequency difference between the uplink and the downlink is 120 MHz.
For each environment, the 3D ray-tracing simulator first generates the channels  between the  user antennas and the corresponding BS antennas\footnote{For more details about how to generate channels, please refer to  \cite{alkhateeb2019deepmimo}.}. Then, the uplink and downlink channels of $U$  selected users are collected as the sample pairs for the corresponding environment\footnote{ {Based on the theory of deep learning, we can improve the generalization capability of the network over the frequency ranges by increasing the randomly selected frequencies, or improve the generalization capability of the network over users' positions by increasing $U$.}}. After obtained the sample pairs for all the environments, we randomly select 1500 environments out as the source  environments and the rest of them are used as the target environments. For each source task, $N_{\textrm{Tr}}$ sample pairs of one source  environment are randomly selected out as the training dataset.
For each target task,  $N_{\textrm{Ad}}$ and $N_{\textrm{Te}}$ sample pairs of one target  environment are separately selected out as the adaption and the testing datasets.
 {Since the perfect channels are not available in practical situation, unless otherwise specified, all the sample pairs in the datasets are estimated by the linear minimum-mean-squared-error (LMMSE) algorithm\footnote{ {In this work, we do not compare the proposed DTL based algorithm with conventional channel estimation algorithms (e.g. least squares, LMMSE and etc \cite{1597555}) since the conventional methods need prohibitively high overheads to achieve the online downlink training and uplink feedback.  While the proposed DTL based algorithms  require neither downlink training nor uplink feedback after the offline training and the adaption stages are finished,  which is a  significant advantage over the conventional methods.  In fact, the best way for the applications of the DTL based algorithms is to serve as complementary approaches of the conventional methods. We can collect adaption data during the applications of the conventional methods and then switch to the DTL algorithms when  the data collection and adaption are finished.}} \cite{1597555} when the signal-to-noise ratio (SNR) is 20 dB and the pilot length is 64. Therefore, the overhead for the  dataset collection is the  pilot length multiplied by twice the number of sample pairs in the datasets.}

The meta-learning, the no-transfer and the direct-transfer algorithms are all implemented on one computer with one GPU.  TensorFlow 1.4.0  is employed as the deep learning framework.
Unless otherwise specified,
the  parameters of the no-transfer, the direct-transfer, and
the meta-learning algorithms are given in Tab.~\ref{tabsfgder}.
The specifical values of these parameters are basically selected by trails and errors such that these algorithms perform well.
 {The FNN adopted in this work has two hidden layers, and each hidden layer has 128 neurons.}
 The numbers of neurons in the input and the output layers are consistent with the lengths of
input and output data vectors, respectively.
The trainable parameters of FNN  are  randomly initialized  as truncated normal variables\footnote{The truncated normal distribution is a normal distribution bounded by two standard deviations from the mean.} with normalized variance\footnote{The  weights of  neurons in the $l$-th layer are initialized as truncated normal  variables with  variance   $1/n_l$, where $n_{l}$ is the number of neurons in the $l$-th layer.}.
To be fair, the number of training samples in $\mathbb{D}_{\textrm{STr}}$ are equal to
\ the number of  training samples used in the meta-learning algorithm, i.e.,   $K_{\textrm{S}}\times N_{\textrm{Tr}}$.
We use the average NMSE as the metric  to evaluate the   performance of algorithms,
 i.e., calculating the average NMSE of the prediction accuracy by repetitively (fine-tuning and then) testing   800 different target environments\footnote{The codes  of this paper are available at  \cite{code}.}.

\begin{table}[!t]\small
\centering
\caption{Default Parameters for the no-transfer, the direct-transfer, and the meta-learning algorithms}
\label{tabsfgder}
\begin{tabular}{c|c}
\hline
Parameter  & Value \\
\hline
Learning rates:  $(\gamma,\beta)$ & (1e-3, 1e-6) \\
Exponential decay rates for ADAM: $(\rho_1,\rho_2)$ & (0.9, 0.999)\\
Disturbance factor for ADAM: $\varepsilon$  &   1e-08 \\
Number of source tasks: $K_{\textrm{S}}$ & 1500\\
Number of target tasks: $K_{\textrm{T}}$ & 800\\
Number of tasks in each training index: $K_{\rm{B}}$ &    80\\
Number of samples in each source task:  $N_{\textrm{Tr}}$ & 20 \\
Number of samples in the target task: $(N_{\textrm{Ad}},N_{\textrm{Te}})$ & (20, 20) \\
Number of users in each task: $U$ & 25 \\
Number of  gradsteps: $(G_{\textrm{Tr}},G_{\textrm{Ad}})$ & (3, 1e3) \\
Batch size: $V$ & 128 \\
\hline
\end{tabular}
\end{table}

 {\subsection{Complexity analysis}
Tab.~\ref{tabsee} compares the complexities of the no-transfer, the direct-transfer, and the meta-learning algorithms.
In the training or adaption stages, the three algorithms repeat the following processes until the loss functions converge: 1) conduct forward propagation of the input and obtain the loss;
2) obtain the   derivatives of the loss with respect to the network parameters;
3) update the  network parameters. Since the  derivative process requires  much more time than the other two processes, we mainly focus on the complexities involved in derivatives.
Based on  \cite{Alex22}, the meta-learning algorithm requires $(G_{\textrm{Tr}}+1)$-order derivatives, and therefore has a higher complexity than the other two algorithms in the training stage.
 Tab.~\ref{tabsee} displays the time required for the three algorithms to
complete the training stage when $G_{\textrm{Tr}}$ is 3.
In the adaption stage, the meta-learning algorithm  requires 1-order derivatives like
the  direct-transfer algorithm since across-task update is no longer needed. Tab.~\ref{tabsee} shows the time required for the networks to adapt to a new environment when $N_{\textrm{Ad}}$ is 20.
It should be mentioned that once the adaption stage is finished, the network would be applicable for all users in the new environment.
Given that the area for one environment is 20$\times$20 m$^2$ in our simulations,
 the proposed algorithms are not competitive for high-speed users (e.g., users with the movement speed higher than 40 m/s). It should be mentioned that the max speed  of users can be increased by deploying   a higher computational power hardware in BSs or  selecting a lager learning rate for adaption.
%
In the testing stage, the three algorithms only need to conduct a single-forward propagation of the input.
Since the three algorithms all adopt the same FNN architecture, the total number of floating point operations for all the three algorithms is $\sum_{l=1}^{L}n_{l-1}n_{l}$.
Tab.~\ref{tabsee} displays the time required for the three algorithms to
complete a single-forward propagation of one input vector.}

\begin{table*}[!t]\small
\centering
\caption{ {Complexity analysis and time cost for the no-transfer, the direct-transfer, and the meta-learning algorithms.}}
\label{tabsee}
\begin{tabular}{c|c|c|c}
\hline
Algorithm  & Training stage  & adaption stage  & testing  stage\\
\hline
No-transfer &  1-order  (17.2 s) & - & $\sum_{l=1}^{L}n_{l-1}n_{l}$ (1.3e-6 s) \\
Direct-transfer &  1-order  (17.2 s)  &  1-order  (0.25 s) & $\sum_{l=1}^{L}n_{l-1}n_{l}$ (1.3e-6 s)  \\
Meta-learning  &  $(G_{\textrm{Tr}}+1)$-order  (65.7 s) &  1-order  (0.24 s) & $\sum_{l=1}^{L}n_{l-1}n_{l}$ (1.3e-6 s)  \\
\hline
\end{tabular}
\end{table*}

\subsection{Performance Evaluation}\label{secres}
\begin{figure}[!t]
\centering 
\begin{minipage}[b]{0.48\textwidth} 
\centering 
\includegraphics[width=1\textwidth]{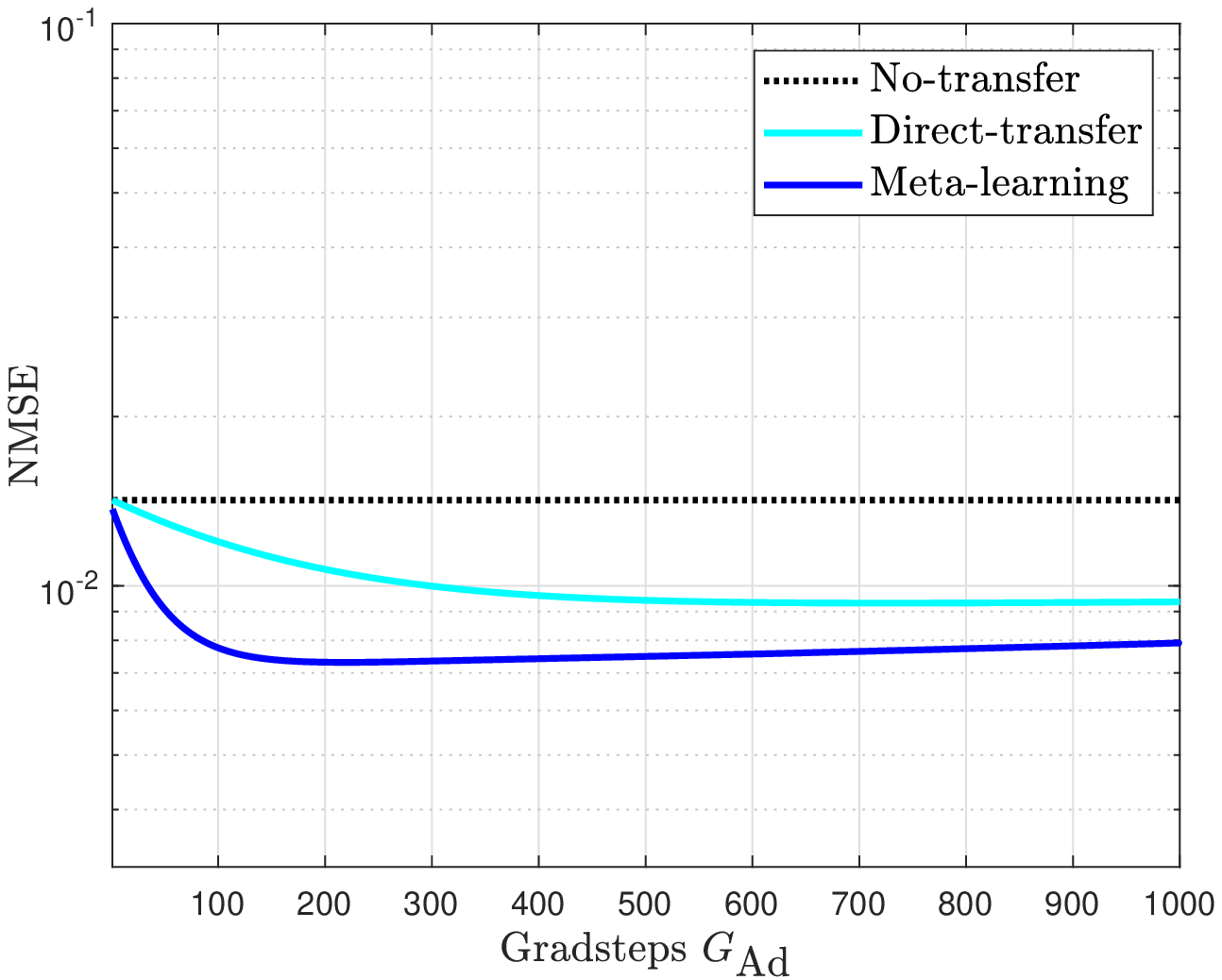} 
\caption{The NMSE performance of  the no-transfer, the direct-transfer, and the meta-learning algorithms versus the number of  gradsteps $G_{\textrm{Ad}}$.}
\label{figsgradstep}
\end{minipage}
\begin{minipage}[b]{0.48\textwidth} 
\centering 
\includegraphics[width=1\textwidth]{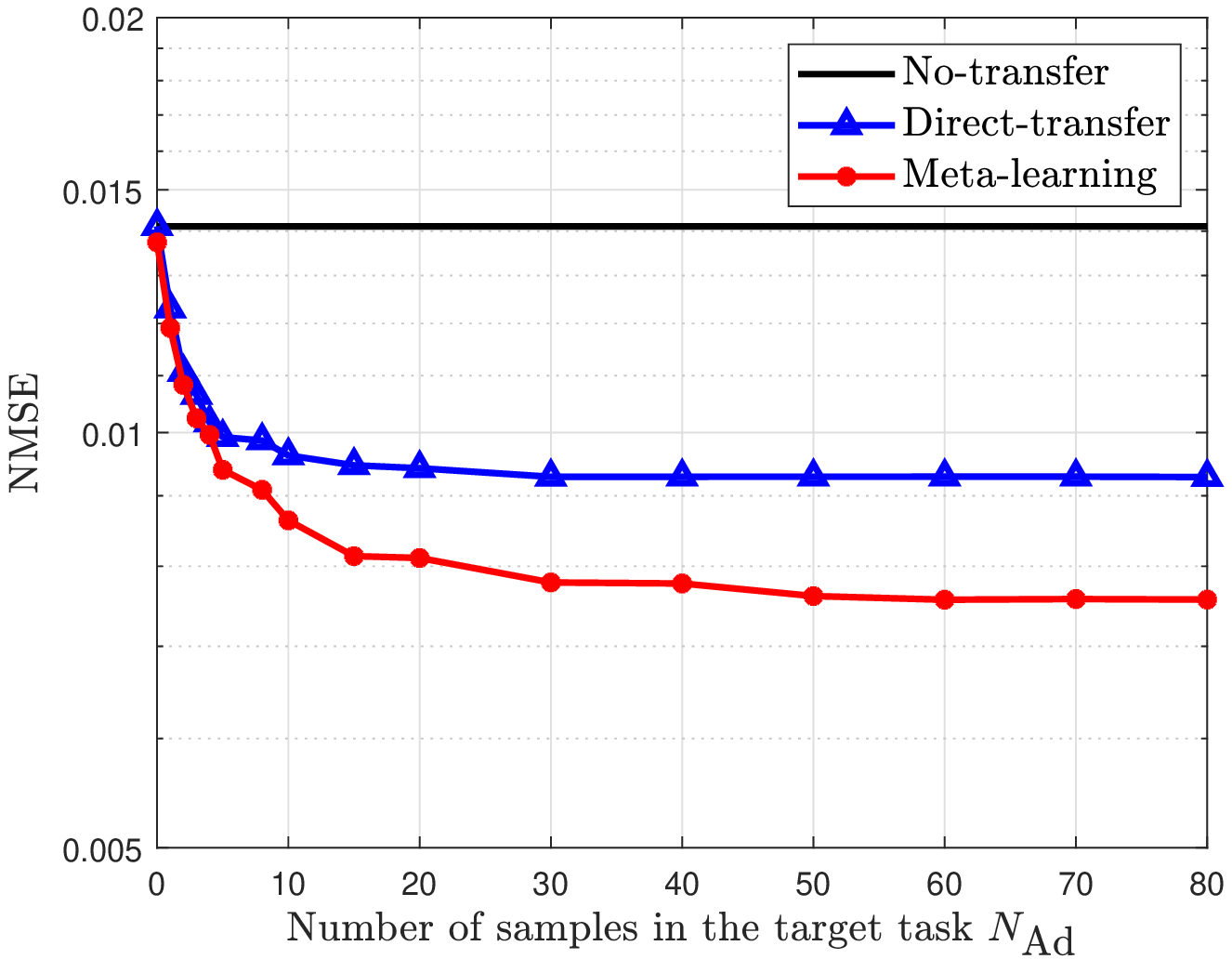}
\caption{The NMSE performance of  the no-transfer, the direct-transfer, and the meta-learning algorithms versus  the number of samples in   $\mathbb{D}_{\textrm{AdSup}}$.}
\label{figsample}
\end{minipage}
\end{figure}
Fig.~\ref{figsgradstep} depicts the NMSE performance of  the no-transfer, the direct-transfer, and the meta-learning algorithms (i.e., $ \rm{NMSE}_{\textrm{Nt}}$, $ \rm{NMSE}_{\textrm{Dt}}$ and $ \rm{NMSE}_{\textrm{Mt}}$) versus the number of  gradsteps $G_{\textrm{Ad}}$.
 Since the network parameters of the no-transfer algorithm do not fine-tune based on the target environment, the accuracy curve  of the  no-transfer algorithm is a horizontal line.
 As shown in  Fig.~\ref{figsgradstep},
 both  the  meta-learning  and the direct-transfer algorithms   are significantly better than the no-transfer algorithm, which indicates that additional samples for the target task and appropriate fine-tuning  can improve the prediction accuracy, i.e., transfer learning  is more suitable than classical deep learning in downlink channel prediction.
Furthermore, the  meta-learning algorithm outperforms the no-transfer and the direct-transfer algorithms, which  demonstrates the superiority of the meta-learning algorithm.
  The performance of  the meta-learning algorithm without meta-adaption stage is similar with  the no-transfer algorithm, however, the NMSE of the meta-learning algorithm drops much  faster than the no-transfer  algorithm as the gradsteps  $G_{\textrm{Ad}}$ increases, which demonstrates that the meta-learning algorithm can find a better initialization for fast adaption than the  direct-transfer  algorithm.
 { We have conducted extensive simulations to investigate the impacts of learning rate $\beta$ on the performance of the direct-transfer and the meta-learning algorithms in the adaption stage.
Simulations show that a large learning rate can speed up the convergence but would cause oscillations and overfitting. In practical systems, we can select a lager learning rate for a faster convergence, select a smaller learning rate for a more stable convergence, or adaptively change the learning rate during the adaption stages to achieve a better balance.
It should be emphasized that no matter what value $\beta$ is, the meta-learning algorithm always
achieves a better accuracy than the direct-transfer algorithm.}

\begin{figure}[!t]
\centering 
\begin{minipage}[b]{0.48\textwidth} 
\centering 
\includegraphics[width=1\textwidth]{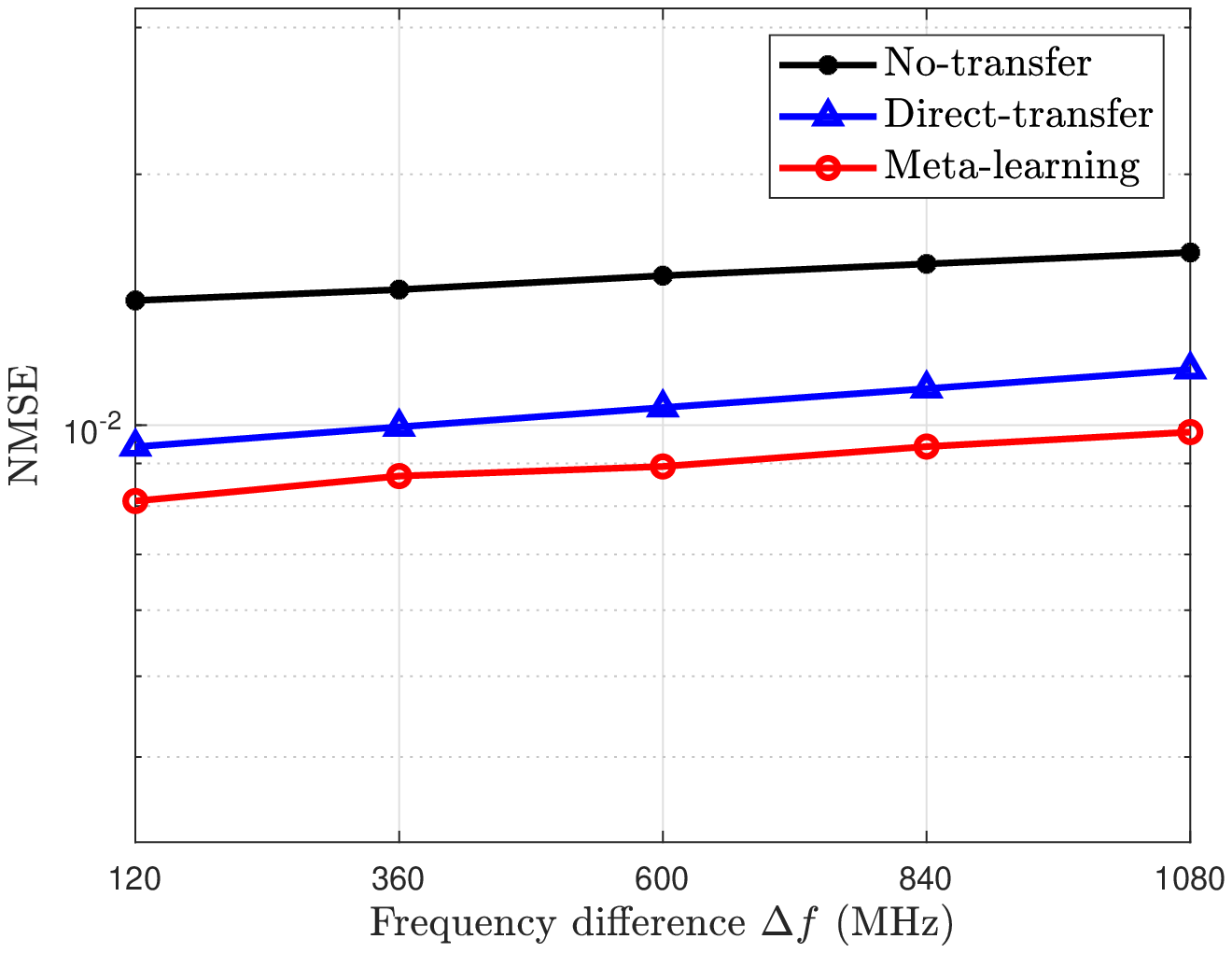} 
\caption{The NMSE performance of the no-transfer, the direct-transfer, and the meta-learning algorithms versus  the frequency difference $\Delta f$.   {The network is trained for each frequency difference $\Delta f$ separately.}}
\label{figfre}
\end{minipage}
\begin{minipage}[b]{0.48\textwidth} 
\centering 
\includegraphics[width=1\textwidth]{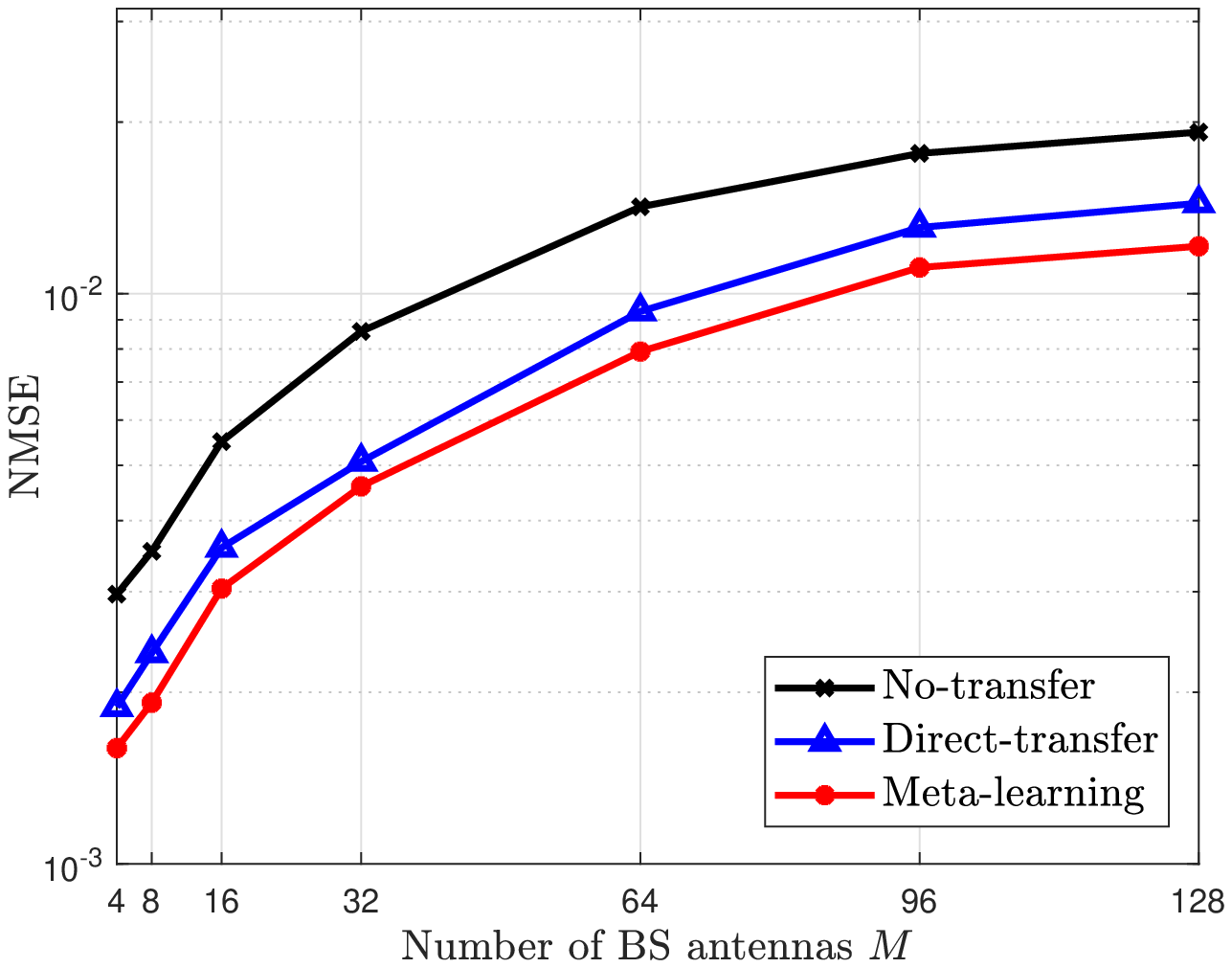}
\caption{The NMSE performance of the no-transfer, the direct-transfer, and the meta-learning algorithms versus  the  number of BS  antennas $M$.   {The network is trained for each BS antenna number separately.}}
\label{figant}
\end{minipage}
\end{figure}

%

Fig.~\ref{figsample} depicts the NMSE performance of
 the no-transfer, the direct-transfer, and the meta-learning algorithms
 versus the number of samples in  $\mathbb{D}_{\textrm{AdSup}}$.
 Since the network parameters of the no-transfer algorithm do not use the dataset $\mathbb{D}_{\textrm{AdSup}}$, the accuracy curve  of the  no-transfer algorithm is a horizontal line.
As shown in Fig.~\ref{figsample}, the meta-learning algorithm consistently outperforms the  direct-transfer algorithm as the sample size increases, which validates its superiority in transfer learning.
Moreover, the prediction accuracies of both the direct-transfer and the meta-learning algorithms  improve as the sample size increases while the improvement rates of both algorithms decrease as the sample size increases.
When the sample size is larger than 60, the performance of the direct-transfer and the meta-learning algorithms both become saturated, which indicates that  a maximum of 60 channel samples in a new environment are required for the network to fully adapt to the new environment, i.e., to predict the downlink channels for any users in the new environment.



Fig.~\ref{figfre} shows the NMSE performance of
 the no-transfer, the direct-transfer, and the meta-learning algorithms
 versus the frequency difference $\Delta f$.
 The prediction accuracies of the three algorithms all degrades as the frequency difference increases.
 This is because  the channel
correlation  between the uplink and the downlink tends to weaken as the frequency difference  increases, and the degradation resulted from the frequency difference is not
destructive for  the uplink based  downlink channel prediction.
The remarkable robustness of DNNs over frequency difference validates the feasibility of channel or beam predictions across  wide bandwidths \cite{alrabeiah2019deep2}.

Fig.~\ref{figant} shows  NMSE performance of the no-transfer, the direct-transfer, and the meta-learning algorithms versus  the  number of BS  antennas  $M$. {For each antenna number point, the adopted FNN has the same hidden layers but different numbers of neurons in the input and the output layers to keep  consistent with the lengths of the input and the output  vectors.}
As shown in Fig.~\ref{figant}, the prediction accuracies of the three algorithms all degrade as the  number of BS antennas increases.  {This involves a common phenomenon ``dimensionality curse'' \cite{bellman1966dynamic}, which refers the phenomenon that when the data dimensionality increases, the dimension of feature space increases so fast that the available data  become sparse and dissimilar in many ways. In this case, the amount of data required to support the data analysis often grows exponentially with the dimensionality. Therefore, the performance of networks often degrades when the data dimensionality increases but the number of training samples does not increase accordingly. When the training dataset is limited, one feasible  solution to improve the network performance is to reduce the lengths of the input and the output vectors. For example, \cite{8052521} proposed to improve the network performance by dividing the input and the output vectors into multiple shorter vectors and then independently training a  network for each of the shortened vector pairs.}
%

\begin{figure}[!t]
\centering 
\includegraphics[width=85 mm]{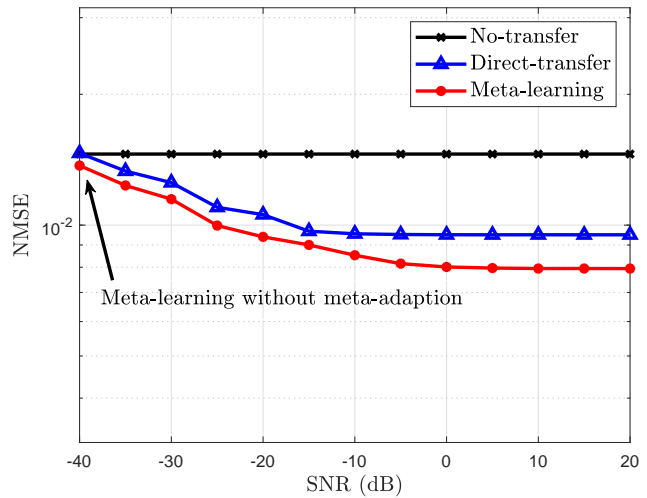}
\caption{ {The NMSE performance of  the no-transfer, the direct-transfer, and the meta-learning algorithms versus  the SNR in collecting adaption datasets.}}
\label{figsnr}       
\end{figure}

 {Fig.~\ref{figsnr} depicts the NMSE performance of  the no-transfer, the direct-transfer, and the meta-learning algorithms versus the SNR in collecting adaption datasets.
The performance of the  direct-transfer and the meta-learning algorithms degrades as SNR decreases when  SNR is lower than 0 dB. When SNR is lower than -40 dB,  the adaption processes cannot improve the performance of the direct-transfer or the meta-learning algorithms.}

\section{Conclusion}\label{secconcul}
In this paper, we   formulated the downlink channel prediction  for FDD massive MIMO systems as a deep transfer learning problem, where each learning task represents   the downlink CSI prediction
from the uplink CSI for a certain environment.
Then, we  proposed the no-transfer, direct-transfer and meta-learning  algorithms based on the
fully-connected neural network  architecture. The no-transfer algorithm trains the network in the   classical deep learning manner.
 The direct-transfer algorithm fine-tunes the network based on the initialization of the no-transfer algorithm. The meta-learning  algorithm learns a model initialization that can effectively adapt
to a new environment with a small amount of labeled data.
 Simulation results have shown that the proposed meta-learning algorithm significantly outperforms the  direct-transfer  and  the no-transfer  algorithms,
 which demonstrates its effectiveness and superiority.


\end{document}